\documentclass{article}
\usepackage{amsmath,amssymb,amsfonts,amsthm,graphicx}
\usepackage[margin=1in]{geometry}
\usepackage{tikz}
\usepackage{hyperref}

\newcommand{\mb}{\mathbf}
\title{Topological Eulerian Synthesis of Slow Motion Periodic Videos\footnote{Supplementary videos can be found at \url{http://www.ctralie.com/Research/SloMoLoops/} } \footnote{ Code to replicate results can be found at \url{https://github.com/ctralie/SloMoLoops} }}

\author{
  Christopher J. Tralie \\
    Duke University,  Department of Mathematics \\
    \texttt{ctralie@alumni.princeton.edu}
   \and
  Matthew Berger \\
	University of Arizona, Department of Computer Science \\
	\texttt{matthew.sh.berger@gmail.com}
}
\date{}
\begin{document}
%
\maketitle

\begin{abstract}

We consider the problem of taking a video that is comprised of multiple periods of repetitive motion, and reordering the frames of the video into a single period, producing a detailed, single cycle video of motion. This problem is challenging, as such videos often contain noise, drift due to camera motion and from cycle to cycle, and irrelevant background motion/occlusions, and these factors can confound the relevant periodic motion we seek in the video. To address these issues in a simple and efficient manner, we introduce a tracking free Eulerian approach for synthesizing a single cycle of motion. Our approach is geometric: we treat each frame as a point in high-dimensional Euclidean space, and analyze the sliding window embedding formed by this sequence of points, which yields samples along a topological loop regardless of the type of periodic motion.  We combine tools from topological data analysis and spectral geometric analysis to estimate the phase of each window, and we exploit the sliding window structure to robustly reorder frames.  We show quantitative results that highlight the robustness of our technique to camera shake, noise, and occlusions, and qualitative results of single-cycle motion synthesis across a variety of scenarios.

\end{abstract}
%
%

\section{Introduction}

Repetitive, periodic behavior is ubiquitous in our world. Animal locomotion, mechanical motion, biological rhythms, and musical rhythms can all be characterized as periodic phenomena.  In this work, we consider the problem of taking a video that contains such repetitive motion of many periods, and synthesizing a single fine detail, slow motion template cycle.  This fine detailed analysis can be applied, for example, to characterize subtle progressions of blood flow in the face during a heartbeat \cite{kumar2015distanceppg}.  Having access to a consensus template can also be used to visualize variations from cycle to cycle. This could be used to indicate the onset of failure in a repetitive automated action on an assembly line, to assess the stress of motions in repetitive human actions~\cite{greene2017visualizing}, and to optimize performance in athletic activities.

Our approach to slow motion templates is {\em Eulerian}; that is, we process the video pixel by pixel with no tracking.  Eulerian approaches for video synthesis are attractive due to their simplicity and ease of implementation.  For instance, a Fourier bandpass filter can successfully elucidate subtle periodic motions in videos \cite{wu2012eulerian, wadhwa2013phase}.  Time-domain Eulerian approaches can synthesize infinitely playing stochastic ``video textures''~\cite{schodl2000video}, as well as seamlessly loop video templates from a single cycle of periodic motion ~\cite{Liao2013VideoLoops,Liao2015VideoLoops}, though such approaches are not applicable to our problem since we fuse multiple cycles.  There are some pitfalls with Eulerian techniques, however.  They are known to fail for large motions, and also require the user to manually specify frequencies of interest \cite{wu2012eulerian, wadhwa2013phase}.  Furthermore, period-to-period drift due to a moving camera or small motion variations can be problematic, as can unrelated motion in the background or occlusions \cite{stauffer1999adaptive}.

\begin{figure*}
\centering
\includegraphics[width=\textwidth]{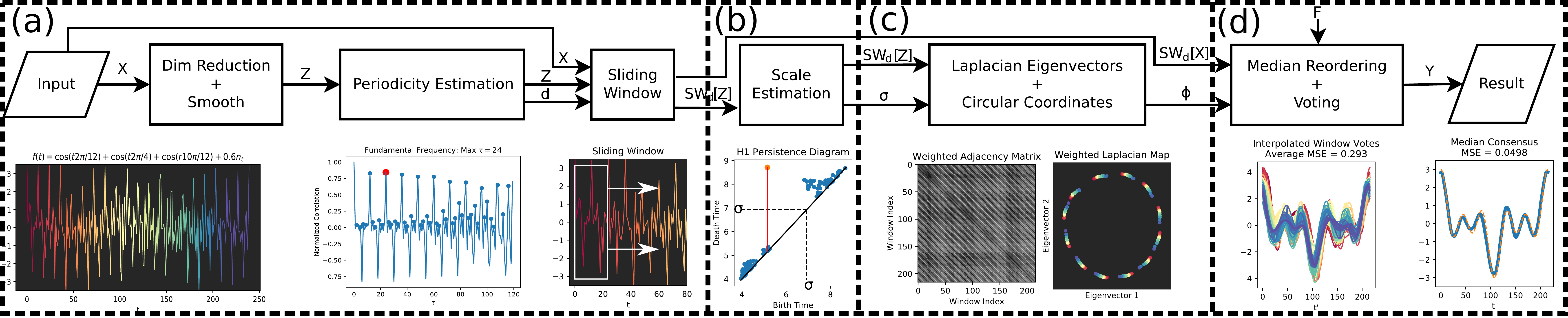}
	\caption{Overview of our technique for a 1D periodic time series with additive Gaussian noise ($n_t \in \mathcal{N}(0, 1)$). We first estimate the signal's period and extract a sliding window embedding (a), followed by estimating the spatial scale (b) via TDA, which is used to compute the phase $\phi$ via the graph Laplacian (c). The single-period motion template is then synthesized by having reordered windows vote on individual pixels (d). On the right we show the result superimposed with ground truth (in orange).}
\label{fig:ConceptFigure}
\end{figure*}

To address the problems with an Eulerian representation, we devise a geometric and topological Eulerian framework by constructing {\em sliding window embeddings} of videos.  Sliding window embeddings, or delay reconstructions, have found a diverse array of applications in activity recognition \cite{frank2010activity,venkataraman2016shape}, video analysis \cite{tralie2017quasi}, and motion capture analysis \cite{venkataraman2016shape}.  Sliding window embeddings of periodic time series, in particular, form samples of a topological loop \cite{perea2015sliding}, as do generalized multivariate sliding windows on periodic video data, regardless of the type of motion present \cite{traliehigh, tralie2017quasi}.  Furthermore, the sliding window embedding of an Eulerian video provides a form of regularization that mitigates the effects of drift~\cite{tralie2017quasi}.  Our geometric framework also allows us to use topological data analysis (TDA) \cite{edelsbrunner2000topological,edelsbrunner2008persistent,edelsbrunner2010computational,carlsson2009topology,ghrist2014elementary} and fundamental frequency estimation \cite{Mcleod05asmarter} to autotune the spatial and temporal scales of the periodic motion, respectively, so that no user intervention is required.  We use Laplacian Eigenmaps~\cite{belkin2003laplacian} computed on the sliding window embedding to obtain a circular phase $\phi \in [0, 2\pi]$ for each window.  We then reorder the windows by $\phi$, and we use windows that overlap in time vote on final pixels in each frame of the template, further mitigating drift.  Figure~\ref{fig:ConceptFigure} highlights the steps of our technique for a sampled 1D periodic time series.  There are only 12 samples per period in the original signal, so the details of each period are coarse and noisy.  However, once resorted, we get a fine-detailed representation of one period.

Many prior works use a combination of sliding window embeddings and TDA, for instance to detect chatter in mechanical systems \cite{khasawneh2016chatter} and quantify periodic activities in motion capture data \cite{vejdemo2015cohomological, venkataraman2016persistent}. However, our approach in using a sliding window jointly with TDA to {\em synthesize} a new result is novel. Related works use spectral geometry to rearrange unstructured collections of images around a loop as a pre-processing step for structure from motion \cite{averbuch2015ringit} or along a line to order the microscopic images of a developing embryo \cite{dsilva2015diffusionvecordering}. By constrast, we seek a cyclic reordering given input frames ordered in time, enabling us to construct a sliding window.


\section{Slow Motion Templates}
\label{sec:methods}


Figure~\ref{fig:ConceptFigure} shows the different parts of our method for a 1D periodic signal for illustration purposes. The actual input to our method is a 3-channel video of $N$ frames, each frame of resolution $W \times H$. We treat each frame as a point in Euclidean space of dimension $\mathbb{R}^{3 \cdot W \cdot H}$, and denote the set of all videos by the sequence $\mb{X}(t) \in \mathbb{R}^{(3 \cdot W \cdot H)}$ for integer-valued $t \in [1,N]$. For steps (a-c), we do not rely on the {\em extrinsic} geometry of the set of points $\mb{X}(t)$, only its set of {\em pairwise distances}. Thus, we can project all of the points into a lower-dimensional space and apply the set of techniques to the projection, assuming that pairwise distances are preserved. Since $N \ll (3 \cdot W \cdot H)$, a distance-preserving projection into an $N-1$-dimensional space is always possible. Denoting this projection by $\mb{P} \in \mathbb{R}^{(N-1) \times (3 \cdot W \cdot H)}$, our method for the steps in (a) operates on $\mb{Z}(t) = \mb{P} \mb{X}(t)$.  We also use the third layer of a Gaussian pyramid on each frame in place of the raw frames before applying $\mb{P}$ to mitigate drift.


\subsection{Sliding Window Embeddings}
\label{sec:slidingwindow}

The sliding window video embedding \cite{cao1998dynamics,traliehigh,tralie2017quasi} for $\mb{Z}$ is defined as
\begin{equation}
	SW_{d}[\mb{Z}(t)] = \left[ \begin{array}{c} \mb{Z}(t) \\ \mb{Z}(t + 1) \\ \vdots \\ \mb{Z}(t + (d-1))  \end{array} \right] \in \mathbb{R}^{(N-1)\cdot(d)},
\end{equation}
where $d$ is the dimension of the embedding. As shown by Takens~\cite{takens1981detecting}, a sliding window of dimension $2m+1$ of even a single generic observation function of a dynamical system of intrinsic dimension $m$ is sufficient to reconstruct a topological embedding of the underlying trajectory in the original state space.  For periodic signals, this state space is a torus, and $d$ should be twice the number of harmonics present to injectively reconstruct loops on that torus \cite{perea2015sliding}.  The same is true of videos in general \cite{tralie2017quasi}.  Furthermore, the sliding window length $(d-1)$ maximizes the topological persistence (Section~\ref{sec:tda}) of the embeddings when $(d-1) = k T$ for some integer $k$, where $T$ is the fundamental period \cite{perea2015sliding, tralie2017quasi}.  To estimate $T$, and hence $d$, we perform 1D ISOMAP \cite{tenenbaum2000global} on $\mb{Z}$ to generate a 1D surrogate signal, and perform an autocorrelation-based fundamental frequency estimation \cite{Mcleod05asmarter}.

\subsection{Laplacian Eigenmaps}
\label{sec:laplacian}

\begin{figure}
\centering
\includegraphics[width=\columnwidth]{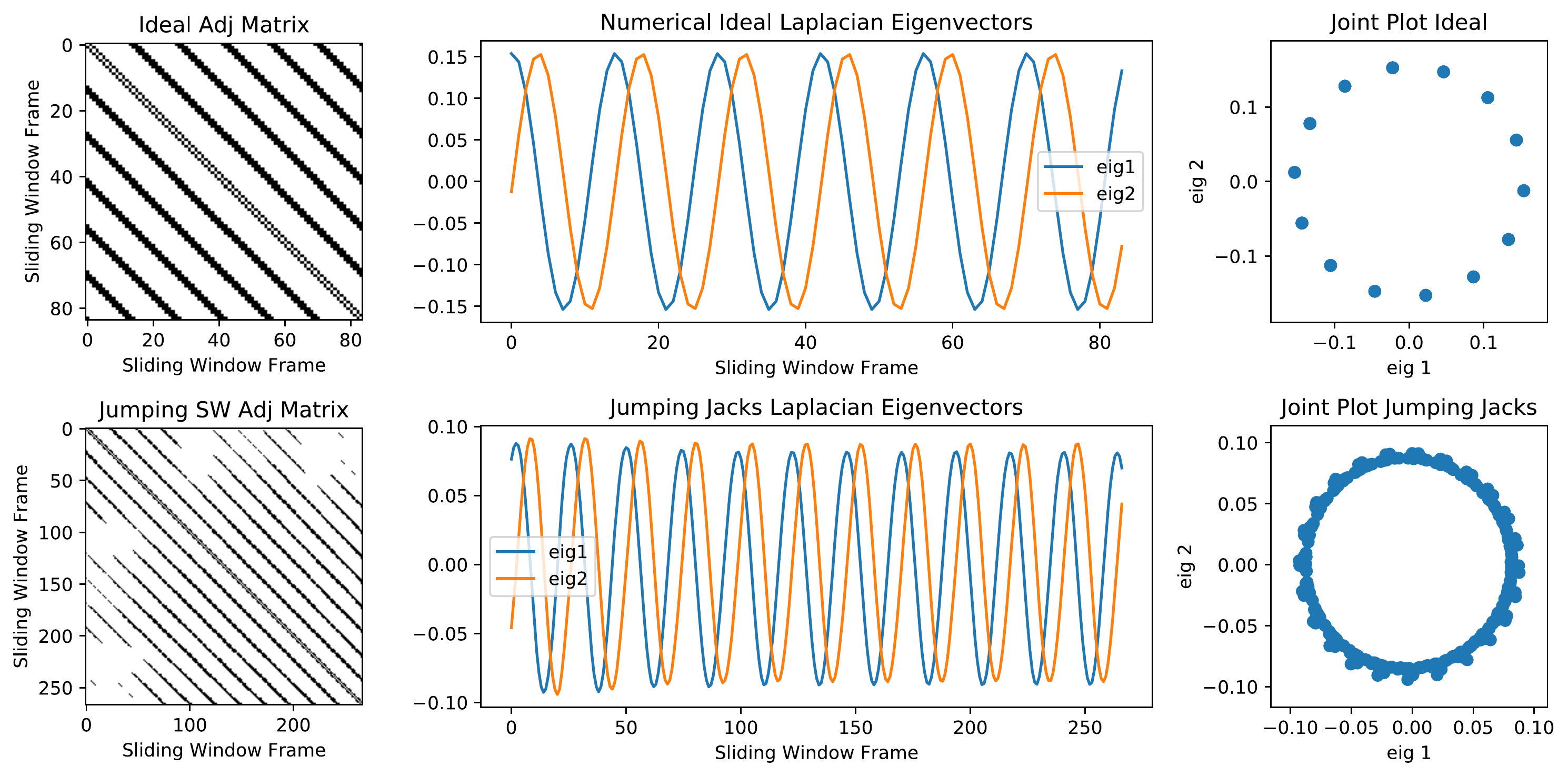}
\caption{An example of the unweighted graph Laplacian on an ideal adjacency matrix corresponding to 6 periods of length 14 (top row), and the adjacency matrix from a sliding window embedding of two men doing jumping jacks (bottom row).  }
\label{fig:CirculantExample}
\end{figure}

Since sliding window embeddings of periodic videos lie on a topological loop, this means that, at an appropriate scale, a graph built on the sliding window point cloud will be approximately circular.  Based on this, we use tools from spectral graph theory~\cite{chung1997spectral}, in particular the {\em graph Laplacian}, to estimate the {\em phase} $\phi \in [0, 2 \pi]$ of each sliding window. We first construct an adjacency matrix $\mb{A}$ in which the entry at $\mb{A}_{ij}$ represents the similarity between windows at times $i$ and $j$.  Given a scale $\sigma$, we define the unweighted adjacency as $\mb{A}_{ij} = 1$ if $||SW_{d}[\mb{Z}(i)] - SW_{d}[\mb{Z}(j)]||_2 \leq \sigma$ and 0 otherwise. We also define a weighted version in which $\mb{A}_{ij} = exp(-||SW_{d}[\mb{Z}(i)] - SW_{d}[\mb{Z}(j)]||_2/2\sigma^2)$. In both cases, we define the graph Laplacian as:
\begin{equation}
	\mb{L} = \mb{D}-\mb{A}
\end{equation}
where $\mb{D}_{ii} = \sum_{j = 1}^N \mb{A}_{ij}, \mb{D}_{i \neq j} = 0$ is the degree matrix representing the sum of all outgoing weights on the diagonal.  

We now examine the unweighted Laplacian of an ideal model to motivate our approach.  Suppose we have extracted a sliding window embedding $SW_{d}$ for a video $\mb{Z}$ which repeats itself exactly every $T$ frames for $k$ times, for a total of $N = kT$ windows.  Let us further suppose it is possible to choose a $\sigma$ so that the unweighted adjacency matrix contains an entry of 1 for edges corresponding to windows that are adjacent in time, as well as for the corresponding windows at, before, and after repetitions of the window at intervals of $T$; that is, the Laplacian is a symmetric, {\em circulant} matrix defined as

\begin{equation}
\label{eq:modellaplacian}
L_{ij} = \left\{ \begin{array}{cc} 3k-1 & i = j \\ -1 & |i-j| = lT, l \in \mathbb{Z}^+ \\ -1 & |i-j| = lT \pm 1, l \in \mathbb{Z}^+ \\ 0 & \text{otherwise} \end{array} \right\}
\end{equation}

Circulant matrices are diagonalized by the Discrete Fourier Transform \cite{godsil2013algebraic}, and their nonzero eigenvalues come in pairs with multiplicity two, with corresponding eigenvectors $v_1[n] = 1$, $v_{2m}[n] = \cos(2 \pi mn / N)$, $v_{2m+1}[n] = \sin(2 \pi m n / N)$, $m \geq 1$.  In the case of Equation~\ref{eq:modellaplacian}, it can be shown using the DFT that the eigenvalues are $\lambda_1 = 0$ and 

\begin{equation}
\begin{array}{cc}\lambda_{2m},\\\lambda_{2m+1}\end{array} = \left\{ \begin{array}{cc} 3k - k\left( 1 + 2 \cos \left( \frac{2 \pi}{kT} m \right) \right) & m = lk, l \in \mathbb{Z}^+ \\ 3k & \text{otherwise}  \end{array} \right\} 
\end{equation}

The smallest two nonzero eigenvalues occur when $m = k$, corresponding to the eigenvectors \footnote{This generalizes the circle graph used in \cite{averbuch2015ringit}, in which $T = N, k = 1$} 
\begin{equation}
v_{2k}[n] = \cos(2 \pi n / T), v_{2k+1}[n] = \sin(2 \pi n / T)
\end{equation}
Therefore, the smallest two numerically nonzero eigenvalues each correspond to sinusoids with period $T$ which are mutually orthogonal.  When plotted against each other, they form a circle with an arbitrary phase offset.  Therefore, we compute the circular phase numerically $\phi = \tan^{-1}(\hat{v_1}[n] / \hat{v_2}[n])$, where $\hat{v_i}$ is the eigenvector corresponding to the $i^\text{th}$ smallest numerically nonzero eigenvalue.

In practice, graphs of the sliding window embedding may deviate from the ideal Laplacian model in Equation~\ref{eq:modellaplacian}, though they do so gracefully.  Figure~\ref{fig:CirculantExample} shows an example of $\hat{v_1}[n]$ and $\hat{v_2}[n]$ for the jumping jacks video.  To improve robustness, we default to the weighted Laplacian so that small changes in the threshold lead to small changes in $\mb{A}$.  In this case, harmonics of the actual frequency of interest occasionally have smaller eigenvalues when the corresponding harmonics are strongly present in the video.  To mitigate this, we search through the 10 eigenvectors corresponding to the smallest 10 eigenvalues, sorted by eigenvalue, and we use the pair of adjacent eigenvectors with the smallest number of zero crossings whose zero crossing counts are within a factor of 20$\%$ of each other.

\subsection{Persistent Homology}
\label{sec:tda}
It remains to find the spatial scale $\sigma$ for the Laplacian.
To adapt to the data, we leverage 1D persistent homology from TDA \cite{edelsbrunner2010computational} to find the scale at which the primary topological feature -- the single cycle -- exists.  Specifically, we compute the 1D Vietoris Rips Filtration on our sliding window data $\mb{Z}$, which tracks equivalence classes of loops, known as {\em homology classes} \cite{Hatcher}.  The algorithm returns a so-called persistence diagram, a multiset of points on a 2D birth/death grid, with each point corresponding to a loop class.  The birth value $b_i$ indicates the scale at which the $i^{\text{th}}$ loop class forms, and the death value $d_i$ indicates the scale at which that class no longer exists.  The difference $d_i - b_i$ is known as the {\em persistence} of the $i^{\text{th}}$ class.  For our scenario, one point in the diagram should have a much larger persistence than the others (e.g. Figure~\ref{fig:ConceptFigure})~\cite{perea2015sliding,tralie2017quasi}, and this ideally reflects the single cycle of motion that we seek.  We take our scale to be $\alpha b_i + (1-\alpha)d_i$ for the largest $d_i - b_i$, where $\alpha \in [0, 1]$ is a parameter which will be explored experimentally in Section~\ref{sec:experiments}.

Finally, homology computation requires the specification of coefficients that belong to a user-given field \cite{Hatcher,maria2014gudhi}. Coefficients in $\mathbb{Z}_2$ are commonly chosen, however in our scenario this is problematic. For instance, the motion of a jumping jack contains a second harmonic, since an individual jumps twice per cycle, and using coefficients in $\mathbb{Z}_2$ would lead to so-called M{\"o}bius splitting~\cite{perea2015sliding, tralie2017geometric, traliemoebius}. Thus, we use coefficients in the field $\mathbb{Z}_{49}$ in order to capture these types of complex motions.

\subsection{Cycle Reordering And Median Voting}
\label{sec:cyclereordering}

Given the phase estimates $\phi_i$, we can now synthesize the final slow motion template by lining up the $N$ sliding windows $SW_{d}[\mb{X}(i)]$ by $\phi$.  For a template with $M$ frames, we choose a set of $M$ equally spaced angles $\theta_1, \theta_2, ..., \theta_M$ around the circle at which to sample the template.  Let $\Phi_i$ be the unwrapped phase of $\phi_i$.  Based on this, we estimate the number of periods $\mb{X}$ goes through as $k = (N \times \max_i(\Phi_i) / (2 \pi (N-d+1)))$.  Assuming $k$ is correct and that $\phi_i$ has a constant slope over $i$, there is a phase gap of $2 \pi k/N$ between adjacent stacked frames in each sliding window.  Thus, $SW_{d}[\mb{X}(i)]$ spans the interval $[\phi_i, \phi_i + 2\pi k d/N (\text{mod} 2 \pi)]$.  A given $\theta_j$ will potentially be contained in the intervals for many different windows, though it is unlikely to coincide exactly with any of the frames in each window, so we use linear interpolation to fill in a frame corresponding to $\theta_j$ in each window overlapping it.  The frame at angle $\phi_j$ is taken as the median of all interpolated frames.
Furthermore, we only use interpolated frames that are a user-prescribed amount $F$ away in time for each given reordered frame, to avoid ghosting artifacts~\cite{meyer2015phase}.

\section{Experiments}
\label{sec:experiments}


\subsection{Quantitative Tests of Circular Coordinates}

\begin{figure}
\centering
\includegraphics[width=\columnwidth]{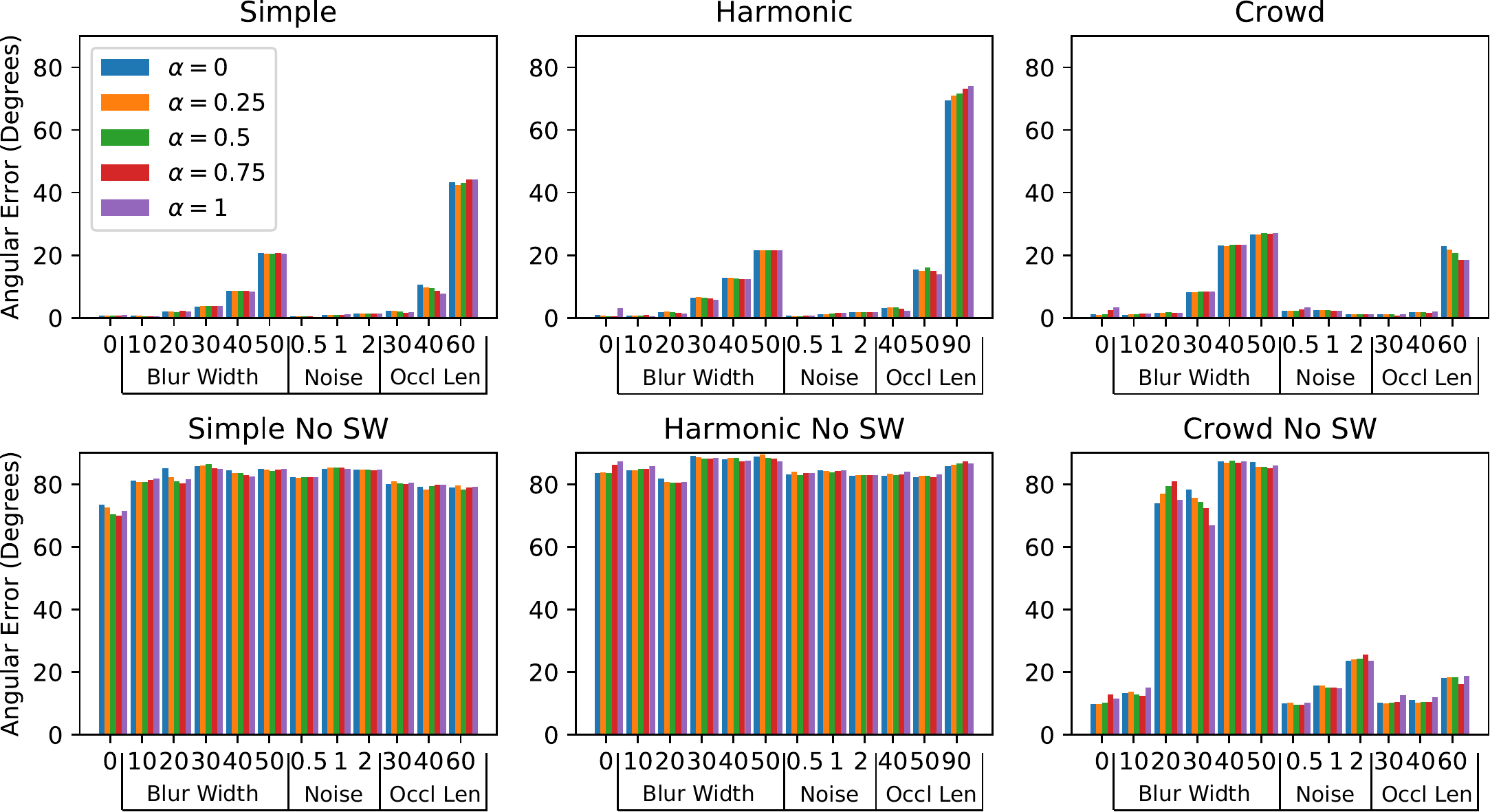}
\caption{Average angular errors of circular coordinates under different sources of corruption on 3 different ground truth videos, with and without a sliding window.}
\label{fig:RobustnessTests}
\end{figure}

We first experimentally quantify the accuracy of our circular coordinate inference, since we cannot hope to get a good slow motion template without accurate circular coordinates.  We generate 3 different 600 frame synthetic periodic videos for which we know the ground truth circular coordinates, using software from \cite{jacobson2012fast}\footnote{Please see supplementary material for these videos and simulated errors.}.  Each video is roughly $300 \times 300$ pixels.  We vary the number of cycles that the videos undergo between $3, 5, 10, 15, 20, 25, 30, 40, 50$.  We then vary the ``shake'' of the video (width of a motion blur kernel) from $10 \times 10 $ to $50 \times 50$ pixels to assess the effect of drift.  We then add Gaussian noise width standard deviations of $0.5, 1, 2$ (original RGB ranges are in $[0, 1]$.  Finally, we add color drifting, occluding squares of varying lengths taking a random walk to assess the effect of occlusions / background motion.  We also compare the $\alpha$ which weights the birth time and death time from TDA.  Figure~\ref{fig:RobustnessTests} shows the average angular error in degrees for our pipeline under these variations, over 50 trials per condition.  Overall, performance is stable to the choice of $\alpha$, and the errors are low for severe noise and for moderate shake and occlusions with a sliding window.  Without a sliding window, the only video that performs reasonably is the ``crowd'' video, though the errors increase more rapidly with shake/noise/occlusion than with the sliding window, validating the ``time regularization'' aspect of sliding windows.

\subsection{Qualitative Video Template Results}
\begin{figure}
\centering
\includegraphics[width=\columnwidth]{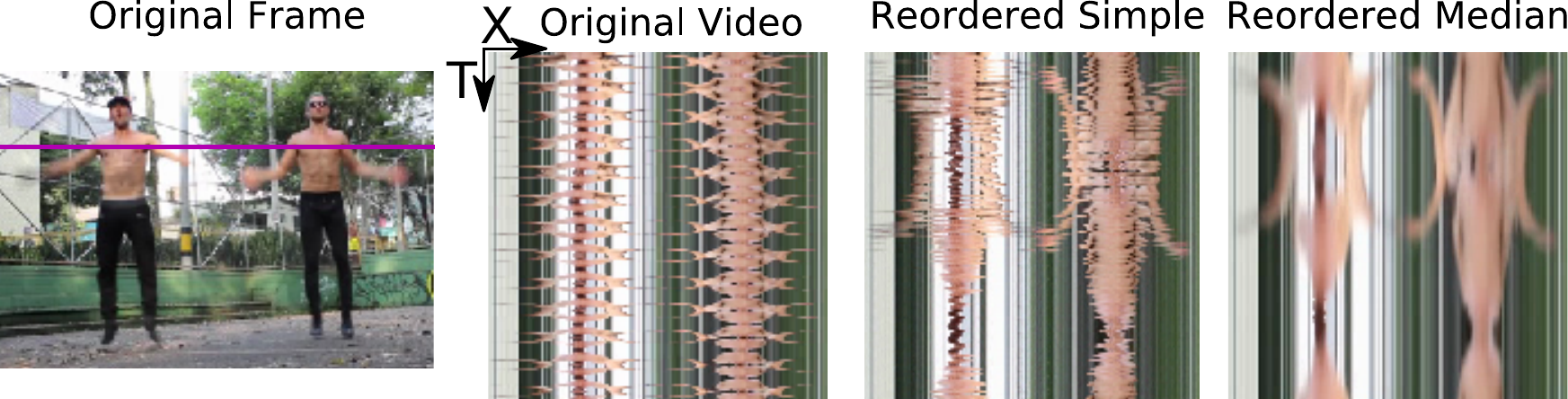}
\caption{An XT slice of a line of pixels (magenta line, upper left) over time for an input video of men doing jumping jacks and for reordered videos with and without median consensus.}
\label{fig:XTSliceJumpingJacks}
\end{figure}

\begin{figure}
\centering
\includegraphics[width=\columnwidth]{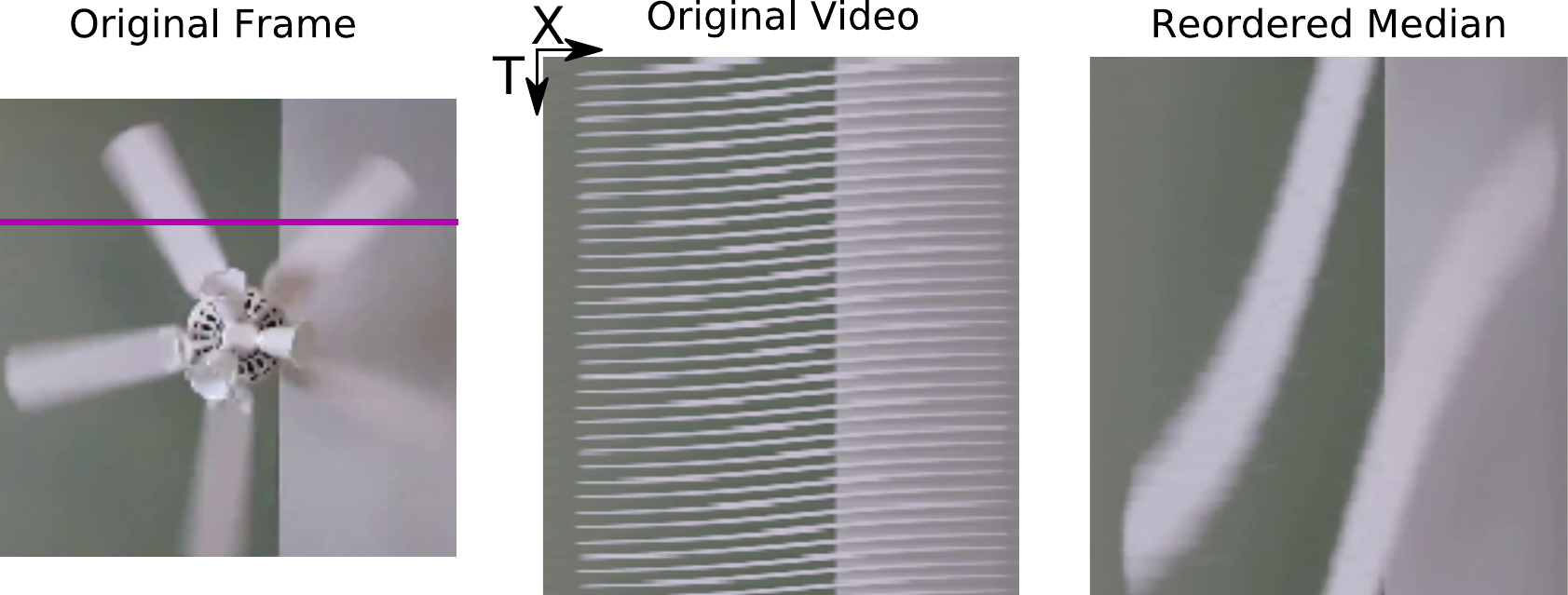}
\caption{An XT slice of a line of pixels (magenta line, upper left) over time for an input video of a fan with only 6 frames per period and the corresponding reordered template.}
\label{fig:XTSliceFan}
\end{figure}

We now qualitatively examine the results of our slow motion templates on some examples.  Figure~\ref{fig:XTSliceJumpingJacks} shows the difference between a simple reordering and a median consensus reordering.  Due to natural variation from cycle to cycle, the simple reordering has many temporal discontinuities when interleaving these cycles.  By contrast, the median voting is clean, and it has the added benefit of removing nonperiodic background components.  Figure~\ref{fig:XTSliceFan} shows an extreme example in which an original spinning fan video has only 6 frames per period at framerate.  Please refer to our supplementary materials for these videos, as well as an exercise video \cite{levy2015live} and videos of amplified blood flow in the neck \cite{wadhwa2013phase} and face \cite{wu2012eulerian}.

\section{Conclusions}

We have presented an approach that combines topological data analysis with spectral geometric analysis to reorder a video consisting of repetitive periodic motion into a single, slow motion template cycle.
Our quantitative results demonstrate robustness to noise, drift, and background outliers, and our qualitative results reveal motion that is challenging to visually perceive from the raw input video.
For future work, we plan on exploring applications of our technique to detecting motion irregularities, visualizing subtle motions from repetitive motion that is temporally aliased, and reconstructing templates for videos with large amounts of missing data.

\bibliographystyle{plain}
\bibliography{refs}

\end{document}